# Plasmon Weyl Degeneracies in Magnetized Plasma

Wenlong Gao[1,2], Biao Yang[1], Mark Lawrence[1], Fengzhou Fang[2], Benjamin Béri[1*], Shuang Zhang[1†]

1. School of Physics & Astronomy, University of Birmingham, B15 2TT, UK
2. State Key Laboratory of Precision Measuring Technology and Instruments, Tianjin University, 300072, China

**Abstract:**

**In this letter, we report the presence of novel type of plasmon Weyl points in a naturally existing material - magnetized plasma. In such a medium, conventional, purely longitudinal bulk plasma oscillations exists only along the direction of applied magnetic field (z direction). With strong enough magnetic field, there exist helical propagating modes along z direction with circular polarizations. The orthogonality between the longitudinal bulk plasmon mode and the transverse helical propagating modes guarantees their crossing at the bulk plasmon frequency. These crossing points, embedded in the bulk plasmon dispersion line, serve as monopoles in the k space – the so called Weyl points. These Weyl points lead to salient observable features, including the striking effect that, at a magnetized plasma surface which is parallel to the applied magnetic field, reflection of an electromagnetic wave with in-plane wave-vector close to the Weyl points exhibits chiral behaviour only in the half k plane bounded by the projection of the bulk plasmon dispersion line. We also verify the presence of 'Fermi Arcs' connecting the two Weyl points with opposite chiralities when magnetized plasma interfaces with trivial photonic materials. Our study introduces the concept of Weyl photonics into homogeneous strongly dispersive photonic materials, which could pave way for realizing new topological photonic devices.**

[*] b.beri@bham.ac.uk
[†] s.zhang@bham.ac.uk

In recent years, studies of photonic topological properties aroused a significant amount of research attention. On one aspect, periodic structures like photonic crystals [1-4], waveguides [5-7], optical resonators[8-10], metamaterials[11-13], and polaritons[14, 15] are studied on their ability to support topological non-trivial edge states mimicking electrons in condensed matter theory. Previous realizations are mainly in two-dimensional (2D) photonic systems, where Dirac points[16, 17] (or quadratic degeneracies which consist of two Dirac points[17]) found at high symmetry points can be described by the Hamiltonian $H = v_x k_x \sigma_x + v_y k_y \sigma_y$ ($\sigma_{x,y,z}$ are Pauli matrices). Opening such Dirac points by breaking time reversal symmetry[1], introducing spin-orbital coupling[6, 11] or synthetic gauge fields[8, 9], one can attain gaps with non-zero Chern numbers or spin Chern numbers that could lead to topologically protected edge states.

The possibility of such gap openings shows that the 2D Dirac points and the corresponding massless excitations are unstable against perturbations[18]. The situation drastically changes in 3D: so-called Weyl points, governed by the Weyl Hamiltonian $H = v_x k_x \sigma_x + v_y k_y \sigma_y + v_z k_z \sigma_z$, provide massless modes that are highly robust [19-21]. A direct way to see this is to note that the inclusion of all three Pauli matrices exhausts all degrees of freedom allowed for a two-band description, making such Weyl-type band crossing protected. More abstractly, Weyl points can be seen as momentum space monopoles of quantized monopole charge, acting as sources of Berry curvature, with the sign of the monopole charge set by the Weyl point's chirality[19]. Within this viewpoint, the stability is understood by noting that the only way to eliminate Weyl points is the annihilation of two anti-chirality Weyl points together. Recently, systems supporting such Weyl points, so-called Weyl semimetals, were observed in electronic[23-26], photonic[27] and acoustic systems[28], which could potentially start a new era in realizing topological states.

So far, in the photonic context, Weyl points have been realised in photonic crystals with precisely engineered lattice structures, periodic on the scale of the wavelength[22]. Here we discover a new type of Weyl points in a homogeneous photonic material: free electron gas under static bias magnetic field (or magnetized plasma). With strong enough applied magnetic field, the cyclotron frequency exceeds the plasma frequency, which results in crossings between the longitudinal plasmon mode and the helical propagating mode at the plasma frequency. These crossing points in the momentum space serve as Weyl points that are responsible for all the non-vanishing Berry curvature and nontrivial topological features in this system. Importantly, magnetized plasma exhibits parabolic equifrequency surfaces (EFSs, the photonic analogue of Fermi surfaces) near the Weyl points which to our knowledge has not yet found a counterpart in condensed matter systems. We also predict salient observable features of these Weyl points in the reflection spectra, including a peculiar polarisation pattern carrying fingerprints of the "eigenstate" properties near the Weyl points. The homogeneity of the system greatly facilitates the investigation of various interesting physics associated with the Weyl degeneracies without relying on complex numerical simulations. In addition, the magnetized plasma approach proposed here does not involve complicated structural design and fabrication, and the system can be reconfigured in real time by varying the plasma density and the strength or direction of the applied static magnetic field.

Unlike artificial metamaterials that could possess highly nonlocal response caused by either nonlocal modes in 'meta-atoms' or Brillouin zone boundary, the magnetized plasma's optical response can be considered to be local as long as the wavelength of interest is far greater than the mean spacing of the charged particles in the plasma. To start, we develop the Hamiltonian formalism of magnetized plasma based on a previous work by Fan et al[29]. The classical equations of motion of a free electron gas under a static magnetic field can be written as:

$$\frac{d^2\vec{P}}{dt^2} - \vec{B}_{static} \times \frac{d}{dt}\vec{P} + \Gamma\frac{d}{dt}\vec{P} = \omega_p^2 \vec{E}, \quad \frac{d}{dt}\vec{P} = \vec{V} \qquad (1)$$

where $\vec{B}_{static} = [0, 0, \omega_c]$ refers to the static bias magnetic field and $\omega_c$ is cyclotron frequency. $\vec{P}$ and $\vec{V}$ are, respectively, the polarization field and electron's velocity field. The damping frequency $\Gamma$ can be neglected as it can be orders of magnitude less than the plasma frequency in a gaseous plasma[31]. By combining Eq. 1 with Maxwell equations and ordering the fields into electro-magnetic fields and electrons velocity fields, we derive the Hamiltonian (relative permeability is 1[30]) as,

$$\begin{bmatrix} 0 & ck\times & -i\mathbb{I} \\ -ck\times & 0 & 0 \\ i\omega_p^2 \mathbb{I} & 0 & \omega_c\Delta \end{bmatrix} \begin{bmatrix} \vec{E} \\ \vec{H} \\ \vec{V} \end{bmatrix} = \omega \begin{bmatrix} \vec{E} \\ \vec{H} \\ \vec{V} \end{bmatrix} \qquad (2)$$

Each element in the matrix is a 3-by-3 matrix, c is the speed of light in vacuum, $\mathbb{I}$ is unit matrix, and $\Delta$ is

$$\Delta = \begin{bmatrix} 0 & -i & 0 \\ i & 0 & 0 \\ 0 & 0 & 0 \end{bmatrix}$$

representing the coupling between $V_x$ and $V_y$ induced by Lorentz force. In Eq. (2) $\omega$, is the eigenvalue of this Hamiltonian. The spectrum contains both positive and negative frequencies that are related by transformations: $\omega \to -\omega$ and $[E, H, V] \to [E^*, -H^*, V^*]$ ('$*$' means complex conjugate) with the momentum fixed. After performing another transformation $[E, H, V] \to [E, H, \omega_p^{-1}V]$, the Hamiltonian Eq. (2) could be reformulated to be hermitian:

$$H' = \omega_p \begin{bmatrix} 0 & k\times/k_p & -i \\ -k\times/k_p & 0 & 0 \\ i & 0 & \omega_c\Delta/\omega_p \end{bmatrix} \qquad (3)$$

where $k_p = \omega_p/c$, and the constant $\omega_p$ will be omitted in the subsequent analysis. Due to rotational symmetry about the magnetic field direction (defined as z in our system), Weyl points, if there are any, are constrained to the [$k_x$=0, $k_y$=0] axis in momentum space. In Fig. 1a, the Weyl degeneracies with $\omega_c = 1.2\omega_p$ are visualized as coloured spots embedded in energy bands against $k_z$ and $k_x(k_y)$. It should be noted that only the positive eigenvalues are shown since negative ones are simply the mirror images of them across the $k_z - k_x$ plane. Across the energy bands, there is a straight line along the $k_z$ axis at $\omega_c = \omega_p$. This is the longitudinal bulk plasmon mode occurring at $\varepsilon_z = 0$, which is induced by the Drude dispersion. It is clearly shown that there are four linear degeneracies along the $k_z$ axis: two between the 1st and the 2nd bands, other two between the 2nd and the 3rd bands. The linear degeneracies are between the bulk plasmon and the circular polarized propagating modes, and are protected, in physical terms, by the orthogonality of the polarisations. Mathematically, as will be made clear later, these linear degeneracies are Weyl points and as such are sources and drains of Berry curvature flux lines. The location of the linear degeneracies in momentum space is expressed by,

$$k_z^{weyl} = \pm\sqrt{\frac{\omega_c}{\omega_c \pm \omega_p}} \qquad (4)$$

Eq. (4) is plotted in Fig.1 (b); note that the outer Weyl points go to infinity when $\omega_c = \omega_p$, and leads to a transition of Berry curvature in infinity. The detailed analysis of this asymptotic behaviour of the Berry curvature is discussed in the supplementary material.

Using the Hamiltonian Eq. (3), the Berry curvature of band n can be computed by $\Omega_n = i\sum_{m \neq n}(\langle n|\partial_{R1}|m\rangle\langle m|\partial_{R2}|n\rangle - (R1 \leftrightarrow R2))/(E_n - E_m)^2$ [33, 34]. It is crucial to stress that the Berry curvatures satisfy spatial inversion symmetry: $\vec{\Omega}(\vec{k}) = \vec{\Omega}(-\vec{k})$ [35] (derivation given in supplementary) and thus the chirality of the Weyl points satisfies the relation $\gamma(k_z) = -\gamma(-k_z)$. Because of this symmetry of the Berry curvature, any closed EFS centred at the origin attain zero total Berry flux, or Chern number. Therefore, we are more interested in the open hyperbolic EFSs embedded in the first band, and thus the outer Weyl points between 1st and 2nd bands. The Berry curvature vector of this band is plotted in Fig. 2(a), showing a source (marked red) and a drain (marked blue) of Berry curvatures in accordance with the fact that chirality γ is 1(−1) for positive (negative) $k_z$. Interestingly, the Berry flux lines are concentrated at the origin of the momentum space. Because of the symmetry relation: $\vec{\Omega}(\vec{k}) = \vec{\Omega}(-\vec{k})$, the origin functions as a source in the upper half k space and a drain in the lower half k space.

To better understand these linear degeneracies, we apply $k \cdot p$ theory[16] (or effective Hamiltonian theory) to obtain the approximate Hamiltonian near the degeneracies (the details can be found in the supplementary). Expanding to first order in the vicinity of a the degeneracy, we find the effective Hamiltonian

$$H_1 = \begin{bmatrix} Mk_z & N_x k_x - iN_y k_y \\ N_x k_x + iN_y k_y & 0 \end{bmatrix} \qquad (5)$$

Here the coordinate origin is shifted to the Weyl point, and $k_{x,y,z}$ represents small deviations from the degeneracy in momentum space. We can re-write Eq.(5) in a more concise form as $H = N_x k_x \sigma_x + N_y k_y \sigma_y + \frac{M}{2} k_z \sigma_z + \frac{M}{2} k_z I$, which is very similar to the expression of Weyl dispersion except for the last term. Under rotational symmetry, $|N_x| = |N_y|$. M and $N_{x,y}$ are real and can be expressed in terms of $\omega_c$ and $\omega_p$. In what follows we focus on the outer Weyl points. They are characterised by:

$$M = \frac{\text{sgn}(\omega_c k_z^{weyl}) 2\sqrt{\varepsilon_{12}}}{\varepsilon_{12}^2 - \varepsilon_{12} + 2}$$

$$N_{x,y} = -\frac{1}{2}\frac{\text{sgn}(\omega_c)\sqrt{\varepsilon_{12}}}{\sqrt{\varepsilon_{12}^2 - \varepsilon_{12} + 2}}$$

where $\varepsilon_{12} = \omega_c/(\omega_c - \omega_p)$. It is worth noticing that when $k_x$ and $k_y$ are zero, the eigenvalues of this Hamiltonian are simply 0 and $Mk_z$. The eigenvalue 0 corresponds to the longitudinal plasma mode, which is a straight line at $\omega = \omega_p$. Interestingly, as is shown in the inset of Fig.1 (b), this Weyl Hamiltonian represents a tilted Weyl cone[31] and is identified to be at the transition between type-Ⅰ Weyl points with a spherical or, more generally, ellipsoid EFSs) and type-II Weyl points with hyperbolic EFSs[32]. Indeed, at frequencies slightly shifted away from the 'Weyl point frequency', the Hamiltonian given by Eq. (5) results in highly anisotropic parabolic EFSs (Fig.1 (c )) expressed by:

$$k_z = \frac{\omega^2 - N_x^2 k_x^2 - N_y^2 k_y^2}{\omega M} \quad (6)$$

The chirality for the Weyl point is defined as $\gamma = sgn(MN_x N_y)$[19-21, 36, 37]. The Berry curvature from this Weyl point can be readily expressed by (see supplementary material):

$$\vec{\Omega}(\lambda_{down}) = -\vec{\Omega}(\lambda_{up}) = \gamma \frac{|\alpha|}{2(\alpha^2 k_z^2 + k_r^2)^{3/2}} \vec{k} \quad (7)$$

where $\lambda_{up}$ and $\lambda_{down}$ stands for upper band and lower band (Fig. 2 (c)) in the effective Hamiltonian, while $\alpha = \frac{M}{2N_{x,y}} = -2/\sqrt{\varepsilon_{12}^2 - \varepsilon_{12} + 2}$ and $k_r = \sqrt{k_x^2 + k_y^2}$. Eqn. 5 represents the Berry flux of a monopole located in an anisotropic k-space with effective permeability given by $[1, 1, \alpha^{-2}]$. Integrating the Berry curvature in Eq. (7) on the EFS described by Eq. (6), one can obtain a value of $2\pi\gamma = \pm 2\pi$ (supplementary), which corresponds to quantized Chern number of $\pm 1$. Intuitively, slope of the parabola is $\pm(2N_x^2 k_x + 2N_y^2 k_y)/\omega M$ and becomes parallel to the z axis at large momentum, thus collecting all the Berry curvature fields. Thus, the effective model not only proves the quantized gauge flux emitted from the Weyl points but also represents a perfect example that an open 'Fermi surface' could attain quantized Chern number. The chirality $\gamma$ of the outer Weyl point on the positive $k_z$ axis is equal to 1 when $\omega_c > 0$. As an example, the Berry curvatures on parabolic EFSs at $\omega_c = 1.2$ are shown in Fig.1 (c); in this case $\alpha = \sqrt{2}/4$ which leads to an anisotropic gauge flux distribution. In the vicinity to the Weyl point, the parabolic EFS is an approximation to the hyperbolic bands that we are interested in. However, at very large k, the parabolic EFS deviates significantly from the actual hyperbolic equifequency surface, whose slope at large k approaches their asymptotes which is finite, thus providing a possible channel from where Berry curvatures of Weyl points can leak into the infinity. This results a non-integer integral of the Berry curvature over the hyperbolic EFS. Nonetheless, it will be shown later that this does not affect the presence of a Fermi arc that connects between the two elliptical EFSs.

Close to the origin of the k space, by applying $k \cdot p$ theory again we can construct a 3-by-3 effective Hamiltonian(containing the first band, its mirror image band in negative frequency regime and a trivial zero energy band), which is expressed as $-i\omega_c k_z k \times$. It is straightforward to work out the analytical expression of Berry curvature of the first band around the origin as (supplementary):

$$\begin{cases} \Omega_k = -\frac{\vec{k}}{|k|^3} (k_z > 0) \\ \Omega_k = \frac{\vec{k}}{|k|^3} (k_z < 0) \end{cases} \quad (8)$$

Eqn. (7) shows that the Berry curvature flux passing through the origin is quantized to unity, which is reminiscent of magnetic flux lines near an infinitesimal vortex in a superconductor. Further numerical calculation confirms that Berry curvature flux lines at $k_z = 0$ are parallel to the $k_x$-$k_y$ plane (Fig.2 (c)), which satisfies the boundary condition of magnetic field at a superconductor surface. Thus, an amusing viewpoint of the $k_x$-$k_y$ plane is that of an infinitely thin "momentum space superconductor" with an infinitesimally small vortex lying at the origin. The discontinuity of the Berry flux lines across the $k_x$-$k_y$ plane indicates the presence of effective "surface currents" on this plane.

We now move to discussing what the observable consequences of these plasmon Weyl points are. One of the most direct probes is provided by optical detection through an angle resolved reflection experiment [27]. In Fig. 3(a) we show the schematic diagram of the experiment, where a high refractive index material with $n_s = 4$ is attached to the magnetized plasma so that the interface is parallel to the applied magnetic field. We consider transverse electric (TE) polarized light shone on the interface with polar angle Θ and azimuthal angle $\psi$ (positive when it lies between positive x, y axis). The location of the outer Weyl point in momentum space could be detected from the reflection spectrum as follows: As is shown in Fig.3 (a), at $\psi = 0$, the incident plane includes the projection of the Weyl point (red spot). A linear degeneracy is clearly observed as the touching point between two bands at $\omega = \omega_p$ in the reflectance spectrum [Fig 3 (b)]. However when $\psi$ is not zero, a gap opens up between the two bands (Fig.3 (c d)). This clearly identifies that the degeneracy occurs at a single point in the k space. Interestingly, EFSs of magnetized plasma can also be retrieved from the reflection intensity. As is shown in Fig. 4(a b), for $\omega_c = 1.2\omega_p$, at shifted frequencies from the plasmon Weyl points, the non-unity reflection regimes corresponds to the EFSs at each operating frequencies.

A remarkable phenomenon associated with the novel plasmon Weyl points is revealed by the polarization state of reflected light for a plane wave at $\omega = \omega_p$ incident at different angles. The Stokes parameter $S_3$ [38] and the polarization states of the reflected waves for both TE and TM incident waves are plotted in Fig. 4 (c d), where arrows indicate the rotation direction of electric fields along with time evolution. For both TE and TM polarizations, the reflected wave's polarizations share the same feature that negligible polarization conversion happens in the negative $k_x$ regime. However the structure becomes quite complicated around the Weyl point when $k_x > 0$. The sudden change occurring at $k_x = 0$ can be explained by the abrupt change of the evanescent eigen-fields induced by the Weyl points. More explicitly, the evanescent 'eigen states' (i.e. with imaginary $k_y$) can be expressed as $[k_x, -\alpha k_z]^T$ for positive $k_x$ regime, while $[0,1]^T$ for the negative $k_x$ [supplementary]. Note that the basis vectors are the two degenerate states at the plasmon Weyl point– a circularly polarized propagating (i.e., helical) mode and a longitudinal bulk plasmon mode. Thus, in the negative half k plane ($k_x<0$), the magnetized plasma shows bulk plasmon like behaviour, whereas in the positive half k plane ($k_x>0$), the bulk plasmon is intermixed with the helical mode to generate a chiral response which leads to a change in the polarization state.

Another important Weyl point feature is the 'Fermi arc' connecting points with opposite chiralities. To investigate these surface states, we study the interface between the magnetized plasma and a topologically trivial photonic material, perfect electric conductor (PEC). As is shown in Fig.5 (b), when $\omega = \omega_p$ there's a hybrid photonic mode (shown in blue curves) connecting two Weyl points as a 'Fermi arc', while the other pair of Weyl points is enclosed by the middle EFS. As is shown in Fig.5 (c), (d), the 'Fermi arc' is robust, and remains present when frequency is shifted from $\omega_p$, even when the EFSs collapse into closed surfaces as in Fig.3 (a). Finally, by applying the approximation near the Weyl points, it can be shown that the 'Fermi arcs' are only present in the positive $k_x$ regime, and approaching the Weyl point with an angle equal to $-\arctan(\sqrt{\frac{\omega_c}{4\omega_p}})$ (Supplementary Information).

In conclusion, we report novel plasmon degeneracies in magnetized plasma, realising the transition between type-Ⅰ and type-Ⅱ Weyl points. We have shown that these plasmon Weyl points lead to a number of striking optical signatures in terms of reflection intensities and polarisation, and result in robust surface "Fermi arcs". With both magnetised plasmas and the required reflectometry readily available, our predictions may pave the way towards experiments in topological photonics with homogeneous dispersive systems and potential applications of surface Fermi arcs in constructing unidirectional waveguides.

Acknowledgement: We thank Mike Gunn for stimulating discussions and feedback. This research was supported by the Royal Society (BB), EPSRC ( SZ) and the Leverhulme Trust (SZ).

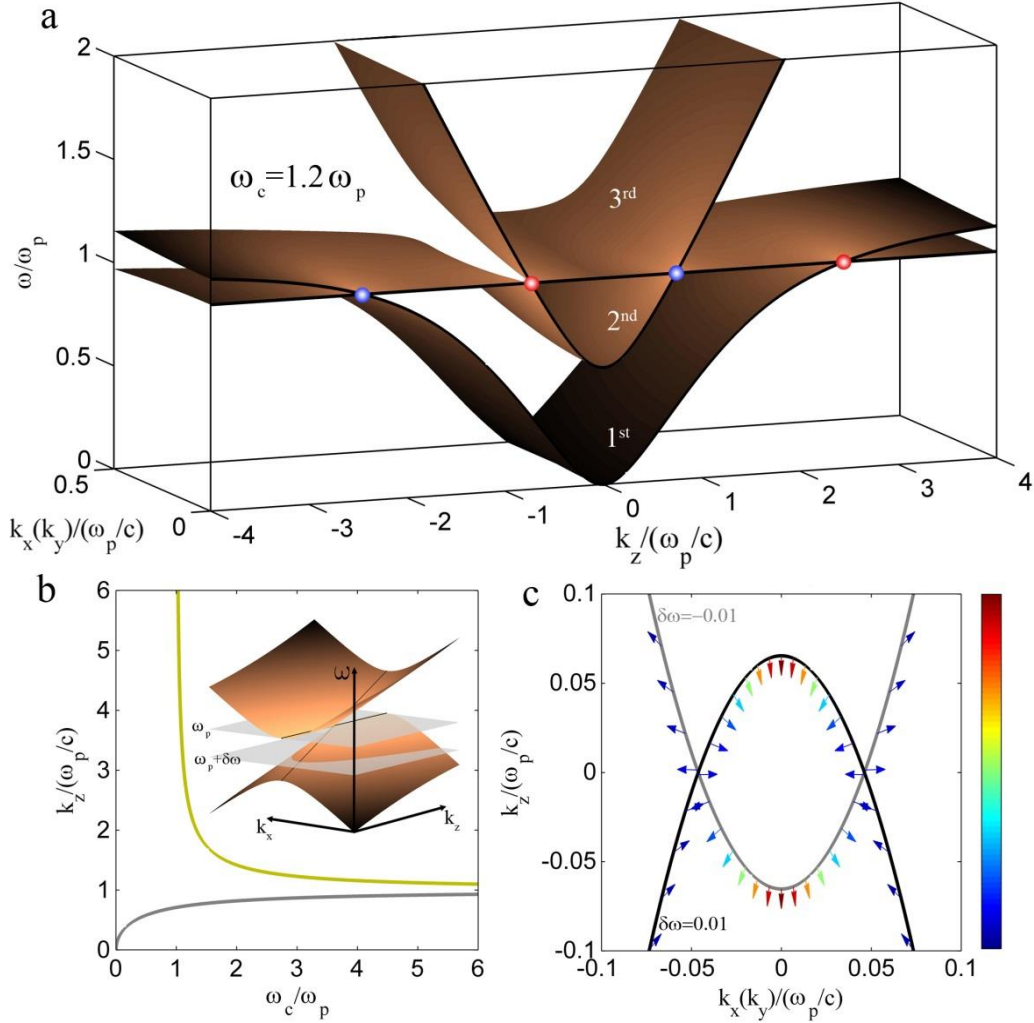

Fig.1 Dispersion relation of a magnetized plasma. (a) 3D energy band of the magnetized plasma for $\omega_c = 1.2\omega_p$, 'Weyl degeneracies' are highlighted by the coloured spots(red for +1 and blue for -1). The straight line at $\omega = \omega_p, k_x = k_y = 0$ is the longitudinal plasma mode. The bands are numbered from bottom to top, and the 4$^{th}$ band is not shown here for conciseness ((b) Plot of Eq. (7) in one quadrant of the parameter space. The outer Weyl points (yellow) goes to infinity when $\omega_c = \omega_p$, while the inner Weyl points (grey) annihilate at the origin. The inset shows the energy bands of the effective Hamiltonian of the plasmon Weyl points. Intersection between the grey planes that pin a definite frequency, and the energy band is a straight line at $\omega = \omega_p$, while is parabola at shifted frequencies (c) 'Fermi surfaces' and Berry curvature plot when shifted frequency $\delta\omega$ equals $-0.01$(grey, corresponds to the 1$^{st}$ band) and 0.01(black, corresponds to the 2$^{nd}$ band ) respectively. The colours indicate intensity of Berry curvatures.

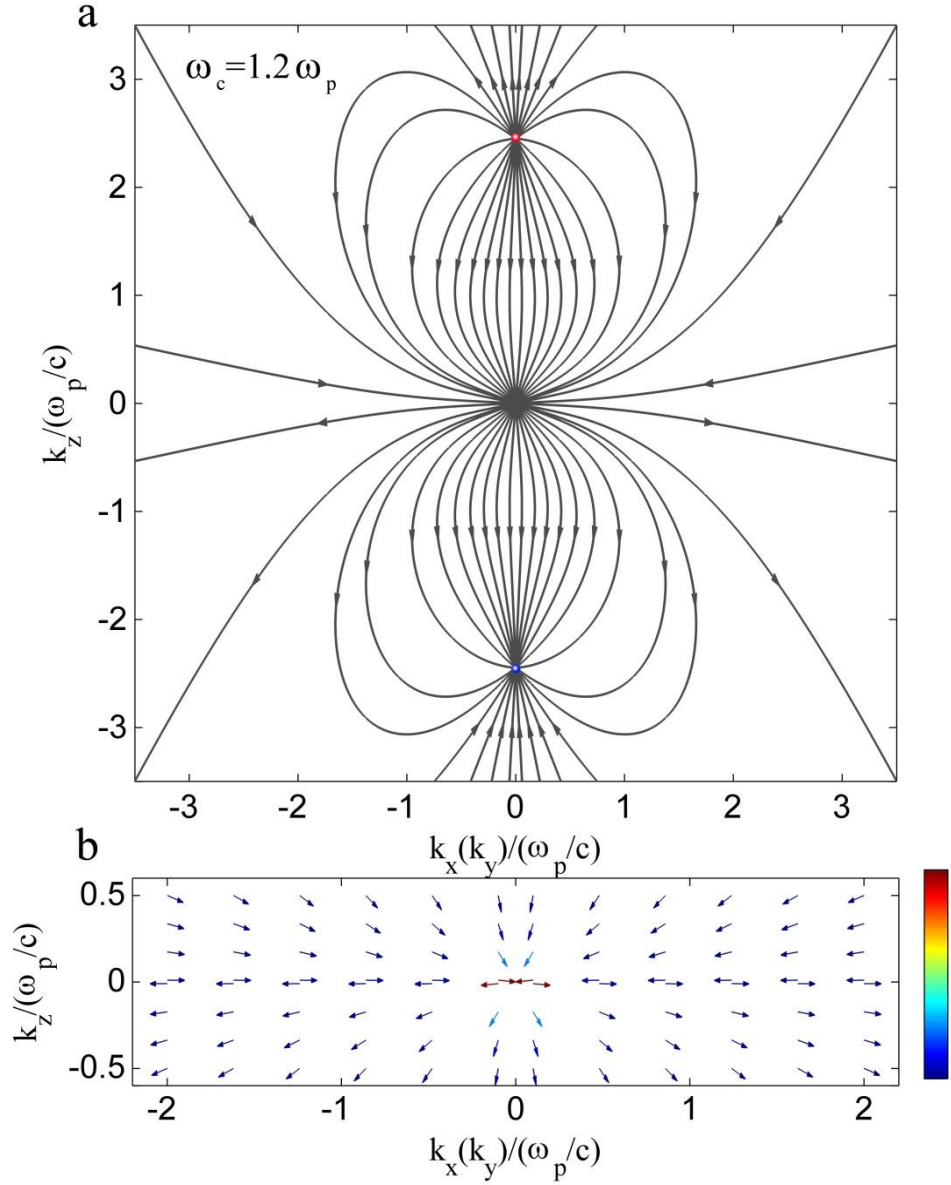

Fig.2 Berry flux distribution of the first band in magnetized plasma. (a) Magnetic field line plot of Berry curvature of the first band. Weyl points are coloured red(blue) for positive(negative) chirality ( (b) Plot of Berry curvatures around the $k_x - k_y$ plane. Colours represent intensity of Berry curvatures, and arrows indicate directions. As is shown, in vicinity to the plane, Berry curvatures are parallel to it.

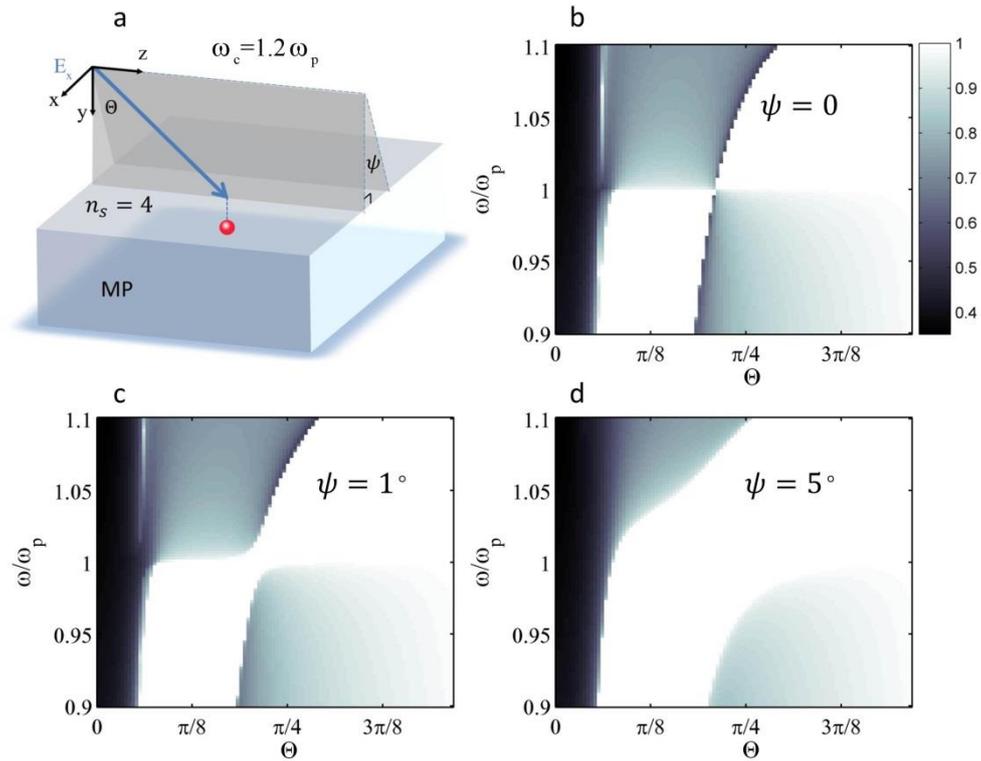

Fig.3 Weyl points probed by reflection spectra of electromagnetic waves by the magnetized plasma. (a) Schematic diagram of the angle resolved reflectivity detection of Weyl point. In this picture, 'Weyl point' is marked as red spot. Note that the azimuthal angle is positive when the incident plane lies between positive x and y axis. (the figure thus shows an angle with negative $\psi$) When the azimuthal angle $\psi$ is zero, the incident plane includes the projection of the Weyl point on the surface. (b) The degeneracy at the Weyl point as seen in the reflection spectrum as the pin point between the two unity reflectivity. (c d) When the azimuthal angle is nonzero, the pin point opens up.

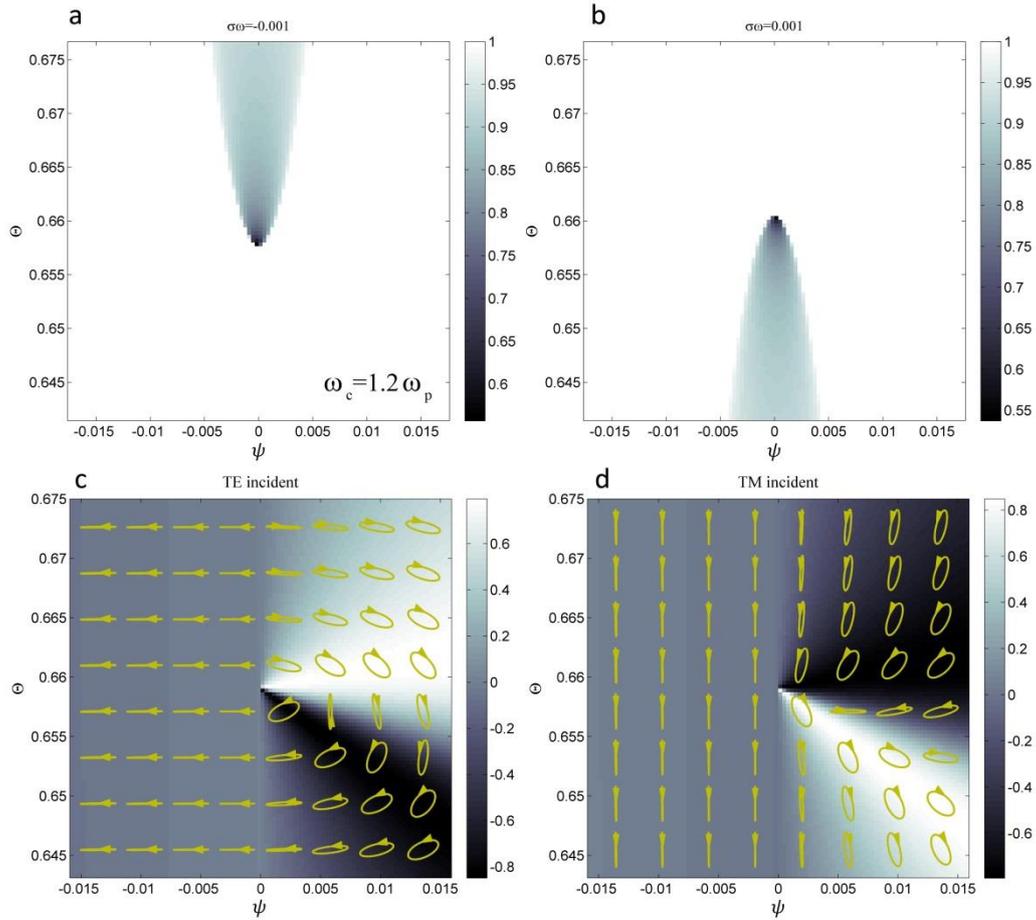

Fig. 4 Intensity and polarization profiles of the reflected waves close to the plasma frequency. (a b) Intensity of reflected light from the angle resolved reflection experiment at operating frequencies shifted from the plasmon Weyl point. Polarizations and $S_3$ parameters (shown as the colour in background) of reflected light for (a) TE polarization and (b) TM polarization around the Weyl points (in the middle of the pictures). As is shown, when incident light is in negative $k_x$ regime, no apparent polarization conversion is observed, while $S_3$ parameters and polarizations change drastically in the positive regime.

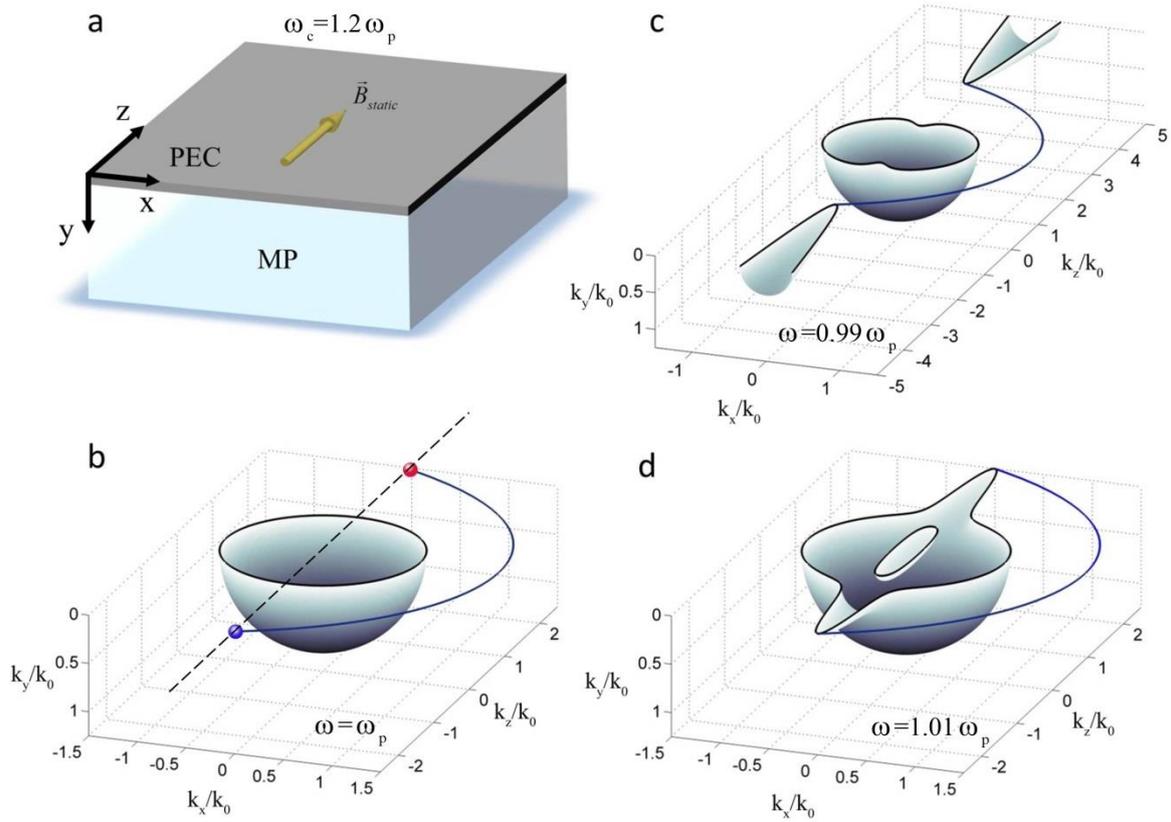

Fig. 5 Presence of a Fermi arc connecting the Weyl points. (a) Schematic diagram of the scheme for realizing 'Fermi arcs' in magnetized plasma. A topologically trivial photonic insulator PEC was attached to bulk Magnetized plasma. (b) At the 'Weyl degeneracy' frequency, there is a 'Fermi arc' connecting the two Weyl points with opposite chirality (marked as red and blue respectively). Away from the Weyl points and the EFS in middle, there is a longitudinal plasma mode (marked as dashed line) along the z axis. (c) The 'Fermi arc' remains persent at shifted frequency equals $0.99\omega_p$. (d) when $\omega = 1.01\omega_p$, there are no open EFSs and total Chern number for the closed EFSs are strictly zero, however the 'Fermi arc' still remains present.